\documentstyle[12pt,procsla,epsfig]{article}
  
  \catcode`\@=11
  \long\def\@makefntext#1{
  \protect\noindent \hbox to 3.2pt {\hskip-.9pt  
  $^{{\ninerm\@thefnmark}}$\hfil}#1\hfill}		
  
  \def\@makefnmark{\hbox to 0pt{$^{\@thefnmark}$\hss}}  
  	
  \def\ps@myheadings{\let\@mkboth\@gobbletwo
  \def\@oddhead{\hbox{}
  \rightmark\hfil\ninerm\thepage}   
  \def\@oddfoot{}\def\@evenhead{\ninerm\thepage\hfil
  \leftmark\hbox{}}\def\@evenfoot{}
  \def\sectionmark##1{}\def\subsectionmark##1{}}
  
  \def\lapprox{\mathrel{\mathop
  {\hbox{\lower0.5ex\hbox{$\sim$}\kern-0.8em\lower-0.7ex\hbox{$<$}}}}}
  \def\gapprox{\mathrel{\mathop
  {\hbox{\lower0.5ex\hbox{$\sim$}\kern-0.8em\lower-0.7ex\hbox{$>$}}}}}

\begin{document}
  
  \centerline{\normalsize\bf Helioseismology, solar models and solar neutrinos
}
  \baselineskip=16pt
  
  \centerline{\footnotesize G. Fiorentini}
  \baselineskip=13pt
  \centerline{\footnotesize\it Dipartimento di Fisica, Universit\'a di Ferrara and INFN-Ferrara}
  \baselineskip=12pt
  \centerline{\footnotesize\it Via Paradiso 12, I-44100 Ferrara, Italy}
  \centerline{\footnotesize E-mail: fiorenti@fe.infn.it}
  \vspace*{0.3cm}
  \centerline{\footnotesize and}
  \vspace*{0.3cm}
  \centerline{\footnotesize B. Ricci}
  \baselineskip=13pt
  \centerline{\footnotesize\it Dipartimento di Fisica, Universit\'a di Ferrara and INFN-Ferrara}
  \baselineskip=12pt
  \centerline{\footnotesize\it Via Paradiso 12, I-44100 Ferrara, Italy}
  \centerline{\footnotesize E-mail: ricci@fe.infn.it}

  \vspace*{0.9cm}
  \abstracts{ We review  recent advances concerning helioseismology,
solar models and solar neutrinos. Particularly we shall address the
following points:
i) helioseismic tests of recent SSMs;
ii)the accuracy of the helioseismic determination of the sound speed near
the solar center;
iii)predictions of neutrino fluxes based on helioseismology, (almost) 
independent of SSMs;
iv)helioseismic tests of exotic solar models.
}

  \normalsize\baselineskip=15pt
  \setcounter{footnote}{0}
  \renewcommand{\thefootnote}{\alph{footnote}}

  \section{Introduction}

Without any doubt, in the last few years
helioseismology has changed 
the perspective of standard solar models (SSM).

Before the advent of helioseismic data 
a solar model had essentially three free parameters
(initial helium and metal abundances, Y$_{in}$ and
Z$_{in}$, and the mixing length coefficient $\alpha$) and produced
 three numbers that could be directly measured:
the present radius, luminosity and heavy element content
of the photosphere.
In itself this was not a big accomplishment and
confidence in the SSMs  actually relied on the success 
of the stellar evoulution theory in describing many and more
complex evolutionary phases in good agreement with
observational data.

Helioseismology has added important data on the solar
structure which provide severe constraint and tests
of SSM calculations. For instance, helioseismology 
accurately determines the depth of the convective zone $R_b$,
the sound speed at the transition radius between the convective
and radiative transfer $c_b$, 
as well as the photospheric helium abundance $Y_{ph}$.
With these additional constraints there are essentially
no free parameters  for SSM builders.

In this paper we review  recent advances concerning helioseismology,
solar models and solar neutrinos. Particularly we shall address the
following points:
i) helioseismic tests of recent SSMs;
ii)the accuracy of the helioseismic determination of the sound speed near
the solar center;
iii)predictions of neutrino fluxes based on helioseismology, (almost) 
independent of SSMs;
iv)helioseismic tests of exotic solar models.

  \begin{figure}
 {\hfill \epsfig{figure=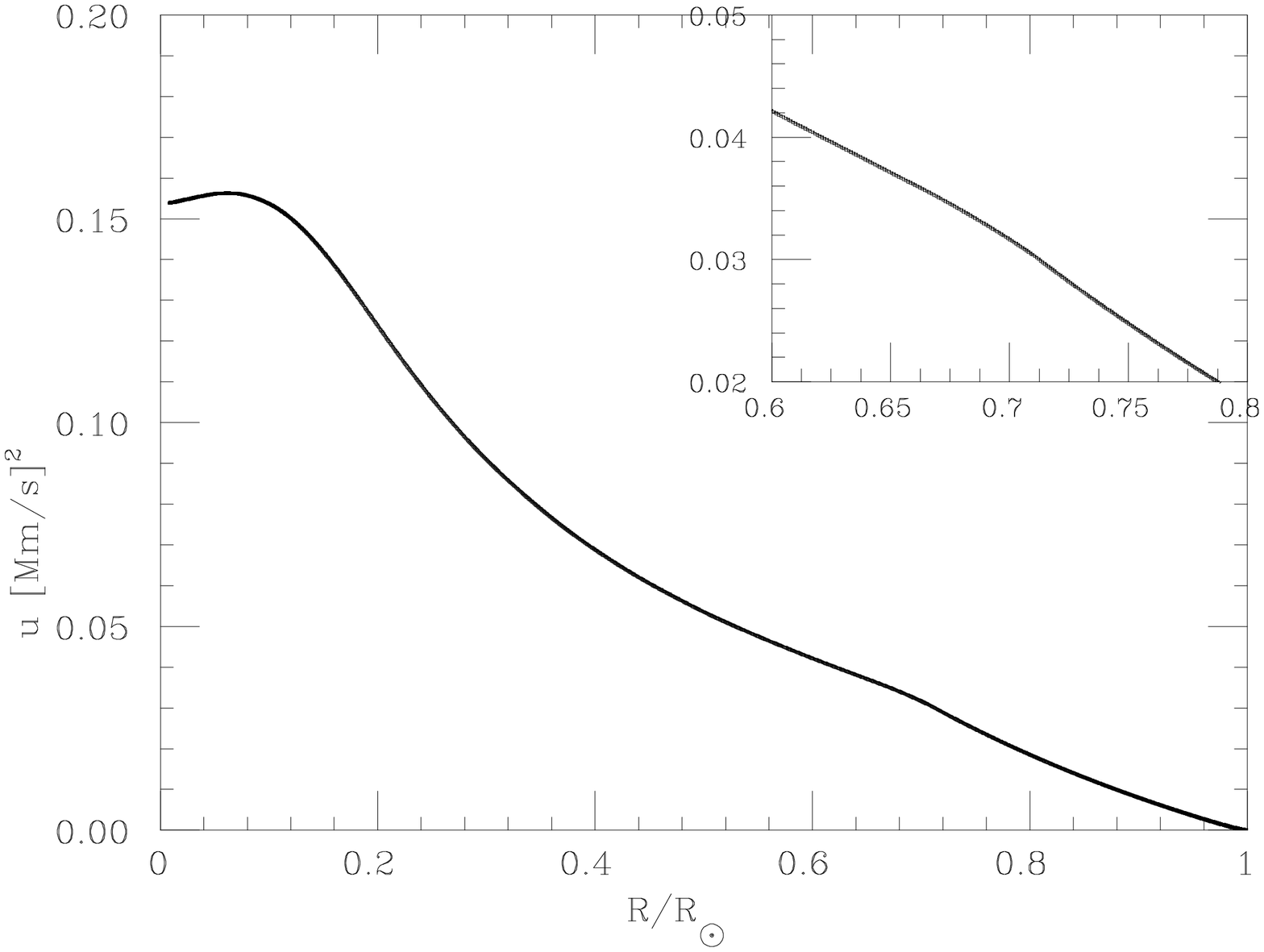,width=10cm,angle=90}\hfill}
  \vspace*{0.2truein}		
  \fcaption{The isothermal sound speed profile, $u=P/\rho$,
as derived from helioseismic observations, from \cite{eliosnoi}.}
  \label{figu}
  \end{figure}

\section{A summary of helioseismic determinations of solar properties}

While we refer to e.g. \cite{eliosnoi} for a review of the
method and to \cite{referenze} for the data, we recall that by measurements of
thousands of solar frequencies (p-modes)with a typical accuracy
of $\Delta \nu/\nu \simeq 10^{-4}$, one derives:

\begin{itemize}

\item
a) properties of the present convective envelope,  
such as depth, density and helium abundance:
\begin{eqnarray}
\label{eqprop1}
R_b&=& 0.711 (1\pm 0.4\%) R_\odot  \\ 
\label{eqprop2}
\rho_b &=& 0.192 (1\pm 0.4\%) g/cm^3 \\
\label{eqprop3}
Y_{ph}&=&x0.249 (1\pm 4\%)
\end{eqnarray}
The quoted errors, mostly resulting from systematic uncertainties in
the inversion technique, have been estimated conservatively by
adding linearly all known individual uncertainties, see \cite{eliosnoi}.
If uncertainties  are added in quadrature, the global error is about
one third  of that indicated 
in eqs. (\ref{eqprop1},\ref{eqprop2},\ref{eqprop3}), 
see again \cite{eliosnoi}.
This latter procedure was also used by Bahcall et al. \cite{bahbasu} 
with similar results. 
This yields the so called  ``$1\sigma$'' errors. We shall refer to the
conservative estimate as the ``$3\sigma$'' determination.
We remark however that this terminology is part of a slang,
and it does not correspond to well defined confidence level,
as one has to combine several essentially systematic errors.

\item
b)sound speed profile. By inversion of helioseismic data  one can determine
the sound speed  in the solar interior. This analysis can be performed
in terms of either the isothermal sound speed squared, $u=P/\rho$, or in terms
of the adiabatic sound speed squared  $c^2= \partial P/ \partial \rho|_{ad}=
\gamma P/\rho$, as the coefficient
$\gamma= \partial log P/ \partial log \rho|_{adiab}$
is extremely well determined by means
of the equation of state of the stellar plasma.

In fig. \ref{figu} we show
the helioseismic value of u as a function of the radial
coordinate $R/R_\odot$. The typical $3\sigma$ errors
are of order $\pm 0.4\%$ in the intermediate solar region,
$R/R_\odot \simeq 0.4$, and increase up to $\pm 2\%$ near the
solar center.

\end{itemize}

\section{Helioseismic tests of  recent Standard Solar Models}

Fig. \ref{figbp98} compares the results of five different observational
determinations of the sound speed in the sun with the results of the
best solar model of ref. \cite {bp98}, hereafter BP98.
This figure suggests several comments:

  \begin{figure}
 {\hfill \epsfig{figure=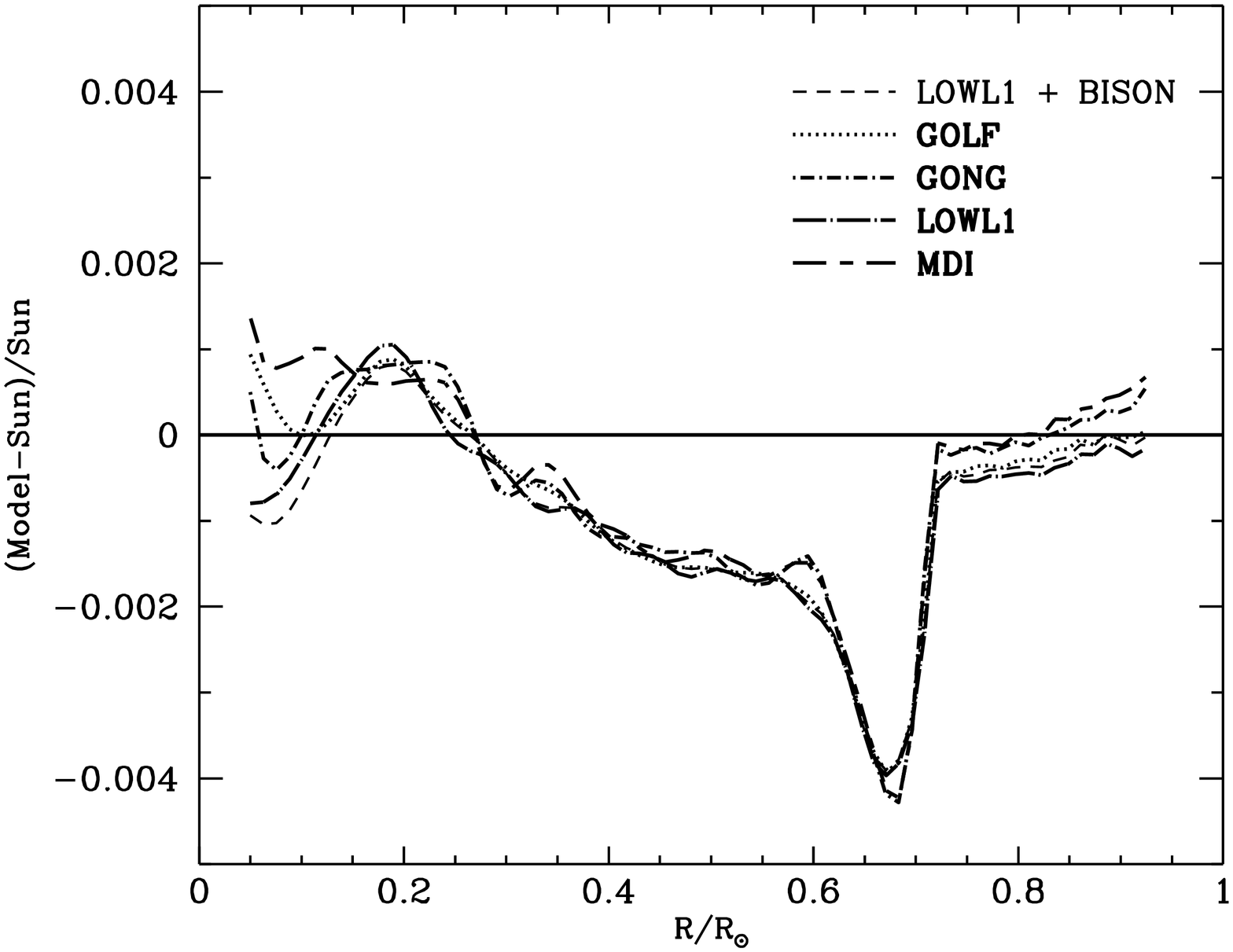,width=10cm,angle=-90}\hfill}
  \vspace*{0.2truein}		
  \fcaption{The predicted BP98 sound speeds compared with five
different helioseismological measurements\cite{referenze}, from \cite{bp98}}
  \label{figbp98}
  \end{figure}

i)Different measurements yield quite consistent value of the sound speed,
to the level 0.1\%;

ii){\it The solar model of BP98 is in agreeement with helioseismic data  to
better than 0.5\% at any depth in the sun}. We remark that
also the properties of the convective envelope predicted
by BP98 ($R_b/R_\odot=0.714,\, \rho_b=0.186 g/cm^3,  \, Y_{ph}=0.243$)
are in good agreement with helioseismic determinations, see eqs.
(\ref{eqprop1},\ref{eqprop2},\ref{eqprop3}).

iii)On the other hand, the predicted sound speed differs  from the
helioseismic determination at the level of 0.3-0.4\%
{\it just below the convective envelope}.

Concerning this last point, we remark that the difference is however
within the ``$3\sigma$''  uncertainty of the helioseismic determination.
Nevertheless it can be taken as an indication of some imperfection of the
SSM. In fact this feature is common to any helioseismic data set, see
again Fig. \ref{figbp98}. As remarked in \cite{eliosnoi}
this feature is common to all recent SSM, which include
elemental diffusion and use updated opacities, see fig. 4 in \cite{eliosnoi}.
It is well known \cite{eliosnoi,bp98}
that by using  older  opacities
the problem disappears or is reduced, so that one can suspect
of the accuracy of the calculated opacity. Furthermore, the well
known  Lithium deficit  
in the photosphere --a factor one hundred below
the meteoric abundance-- in not yet understood.
In addition the ``mixing length theory'' is a  rough description
of the convective transport which maybe a not too good approximation
in the transition between the radiative and the convective region.

In summary  all this means that SSM predictions are  accurate to the level
of one per cent or better, although there are indications of
some deficiencies at the level of per mille.

\section{The solar sound speed in the neutrino production region}

\begin{figure}
 {\hfill \epsfig{figure=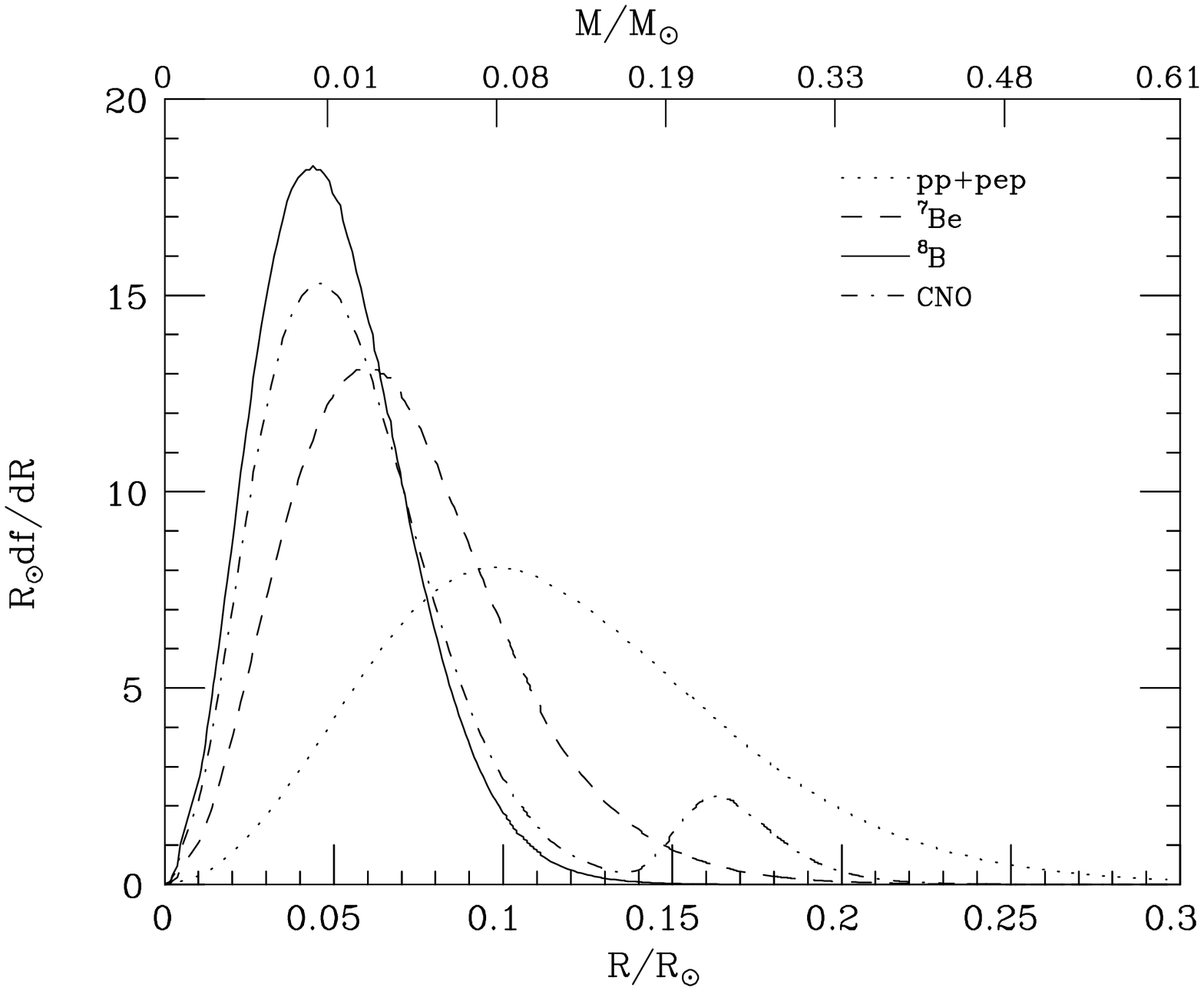,width=10cm,angle=90}\hfill}
  \vspace*{0.2truein}		
  \fcaption{For the indicated components, $df$ is the fraction of neutrinos 
 produced inside the sun within $dR$.
On the bottom (top) scale the radial (mass) coordinate is indicated.}
  \label{figprof}
  \end{figure}

As well known, Boron and Beryllium neutrinos are produced very
near to the solar center, see Fig. \ref{figprof}, with  maximal
production rates respectively at $R_{B}=0.04 R_\odot$ and
$R_{Be}=0.06 R_\odot$. Since the p-modes which are observed do not propagate
(actually are exponentially dumped) so deeply in the sun the question
often arises if present helioseismic data can determine
the sound speed in region of Beryllium and Boron production.

  \begin{figure}
 {\hfill \epsfig{figure=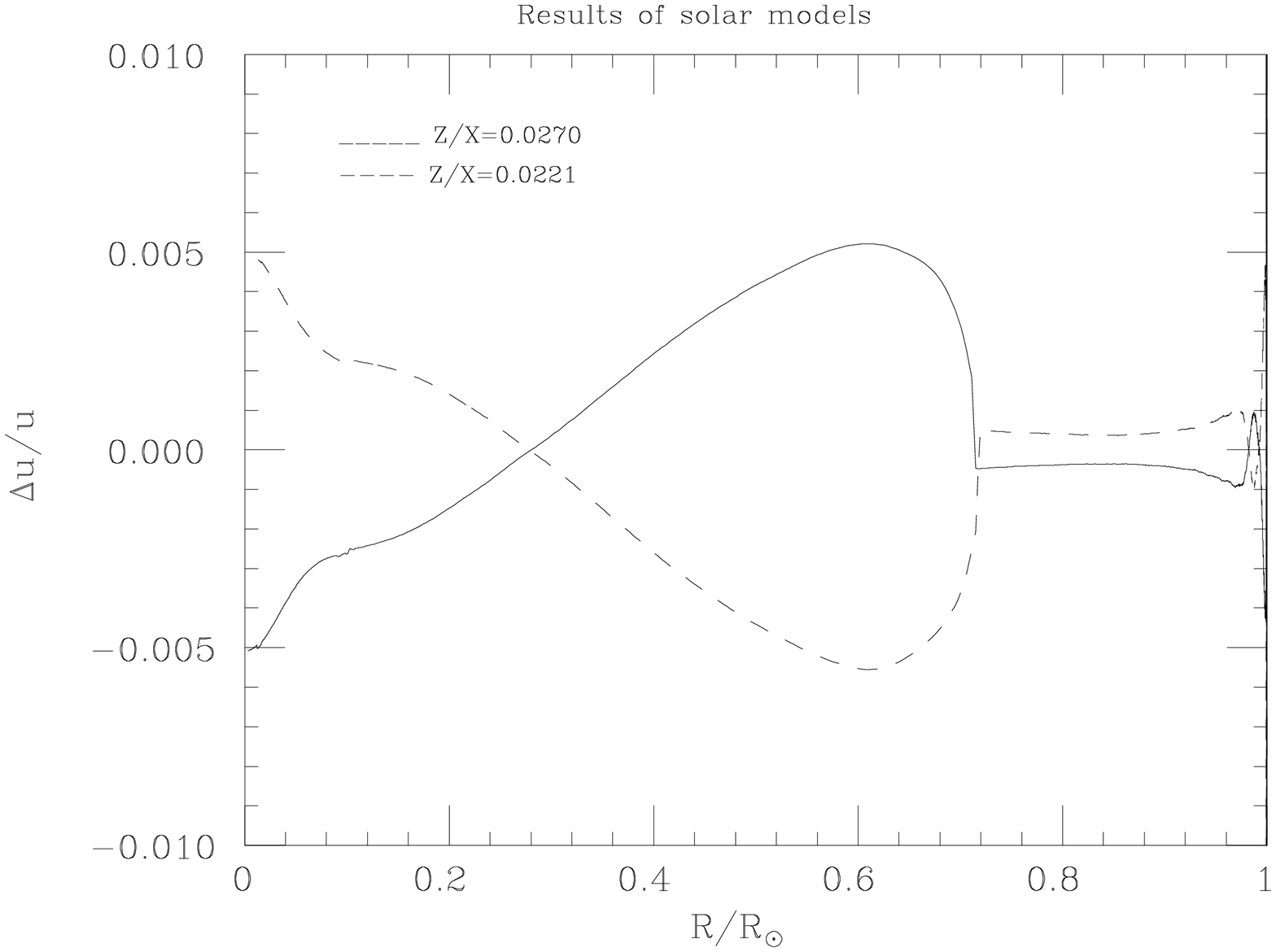,width=10cm,angle=90}\hfill}
  \vspace*{0.2truein}		
  \fcaption{
Difference of the predicted sound speeds of metal rich (poor) models,
compared with respect  to the SSM prediction, full (dashed) line.}
  \label{figmodel}
  \end{figure}

  \begin{figure}
 {\hfill \epsfig{figure=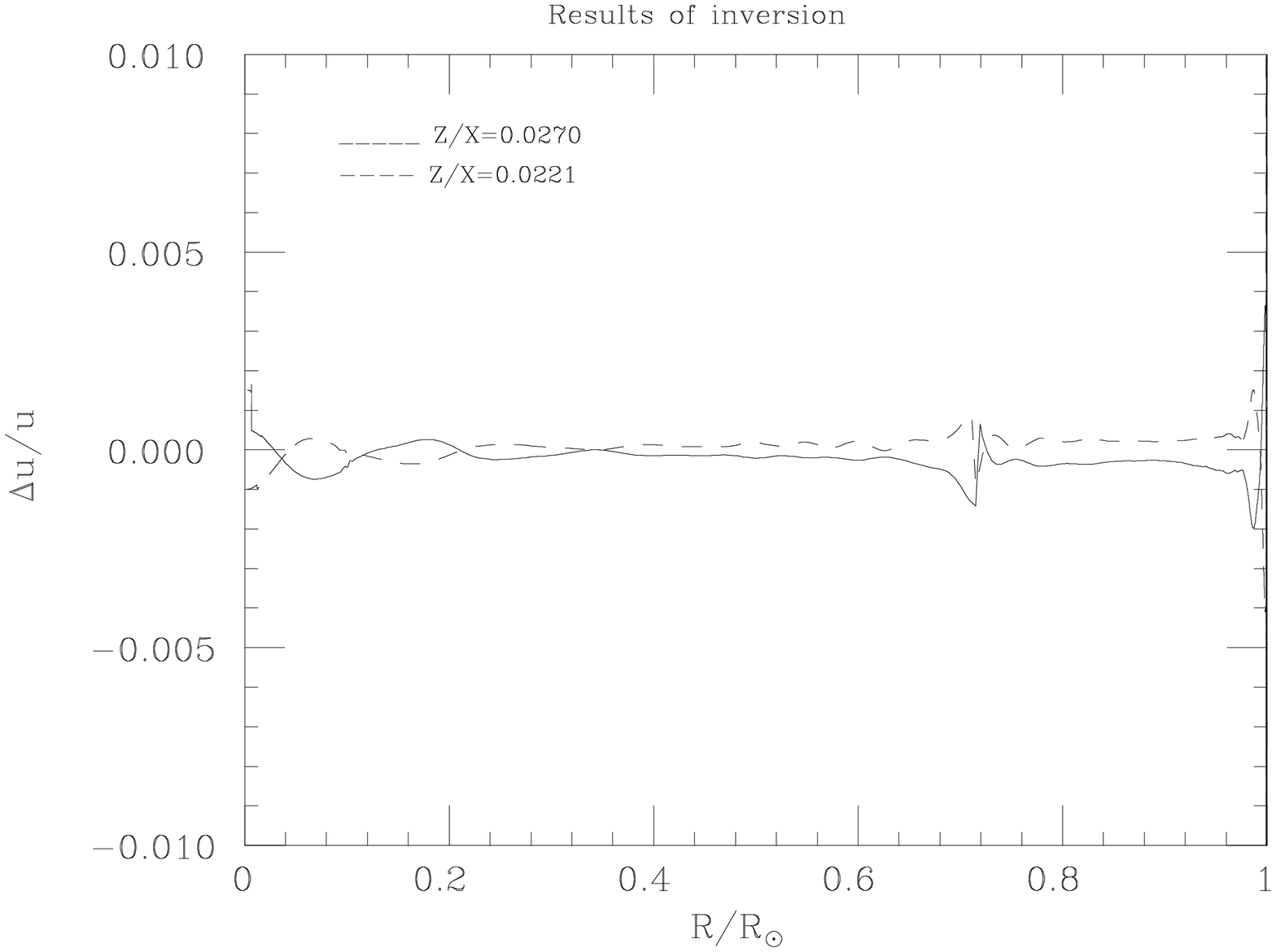,width=10cm,angle=90}\hfill}
  \vspace*{0.2truein}		
  \fcaption{Difference among the helioseismic sound speeds, obtained
by using different starting models.}
  \label{figinv}
  \end{figure}

From an extensive analysis of the inversion method and of data
available at that time we already concluded in \cite{eliosnoi}
that $u(R\simeq0)$ is determined with a ``$1\sigma$'' accuracy of 
1\% In this section we present a simplyfied
analysis in order to produce convincing evidence that helioseismology
fixes the sound speed near the solar center
with such an accuracy.

Essentially there are two questions.

\noindent
a){\it Convergence of the inversion method}: 
for a given helioseismic data set, how does the reconstructed 
sound speed depend on the input solar model? 

\noindent
b){\it Consistency of helioseimic data}: for a given inversion procedure, how 
does the result depend on the helioseismic data set?

In order to address the first question we have used as starting model for
the inversion procedure the SSM of ref. \cite{eliosage} (Z/X=0.0245), 
as well as  two non-standard models:
a metal rich  model (Z/X=0.027) and a metal poor one (Z/X=0.022).
The predicted sound speed differences are shown in Fig. \ref{figmodel}.
We remark that the relative difference of $u$ between 
the metal rich and the metal poor model 
is 1\% at $R\simeq 0$. The helioseismic sound speeds, derived 
starting from these models and by using the 
BBSO86 \cite{BBSO86}+BISON\cite{BISON} data set 
 are shown in Fig. \ref{figinv}. The relative
difference between the reconstructed sound speeds are anywhere 
less or of the order of one per mille.

 In particular {\it the helioseismic sound speeds at 
$R\simeq0$ differ by two per mille although the difference between the models
was a factor 5 larger}. This means that the helioseimic 
sound speed near the center is really determined by data.

The answer to the second question is clearly derived from Fig. \ref{figbp98}.
For a given starting solar model, inversion
of different helioseismic data sets  
gives reconstructed sound speed which differ  as much as two per mille
near the solar center.

In conclusion we confirm our, possibly conservative, ``$1\sigma$'' 
accuracy  
\begin{equation}
\label{deltau}
\Delta u/u (R\simeq 0) =1 \%
\end{equation}

We remind that  what is important for the production of Boron
and Beryllium neutrinos is the solar temperature near
the center. The knowledge of the sound speed does
not give direct information about temperature, 
since the chemical composition and the equation of state
have to be known.

In the energy production zone, the equation of state (EOS) for 
the solar interior can be approximated, with an accuracy
better than 1\%, by the EOS of a fully ionized classical
perfect gas:
\begin{equation}
\label{eqeos}
KT=u \mu \, ,
\end{equation}
where the ``mean molecular weight'' is:
\begin{equation}
\mu= m_p / (3/2X+1/4 Y+1/2) \, .
\end{equation}
For a given value of $u$ without any assumption on the
chemical composition one immediately 
gets a direct helioseismic  constraint on the solar temperature:
\begin{equation}
\label{eqconst}
1/2 u < KT/m_p < 4/3 u \, ,
\end{equation}
which will be useful for the discussion in the next sections.
We remark however that much
more strict bounds on the central temperature can be obtained
by studying the so called Helioseismically Constrained solar
Models HCSM, see \cite{eliostc}.

\section{Predictions on  neutrino fluxes based
on helioseismology}

The basic idea is to use helioseismology in place of 
SSM calculations. Neutrino production rates are generally 
given as:
\begin{equation}
dN/dt= \int dv n_i n_j <\sigma v> _{ij}\propto
     S_{ij} \int dv n_i n_j T^{\alpha_{ij}}
\end{equation}
The astrophysical S-factors $S_{ij}$ are given
by nuclear physics calculations and/or experiments and
the power law coefficients $\alpha_{ij}$ are calculated
by using  the Gamow formula, see. e.g. \cite{rolfs}.
In the usual approach,
the nuclear densities $n_i(R)$ and 
the temperature profile $T(R)$ are given 
by SSM calculations.

Alternatively one can use helioseismology 
to constrain or determine the above integrals.
This can be accomplished in the following way.

  \begin{figure}
 {\hfill \epsfig{figure=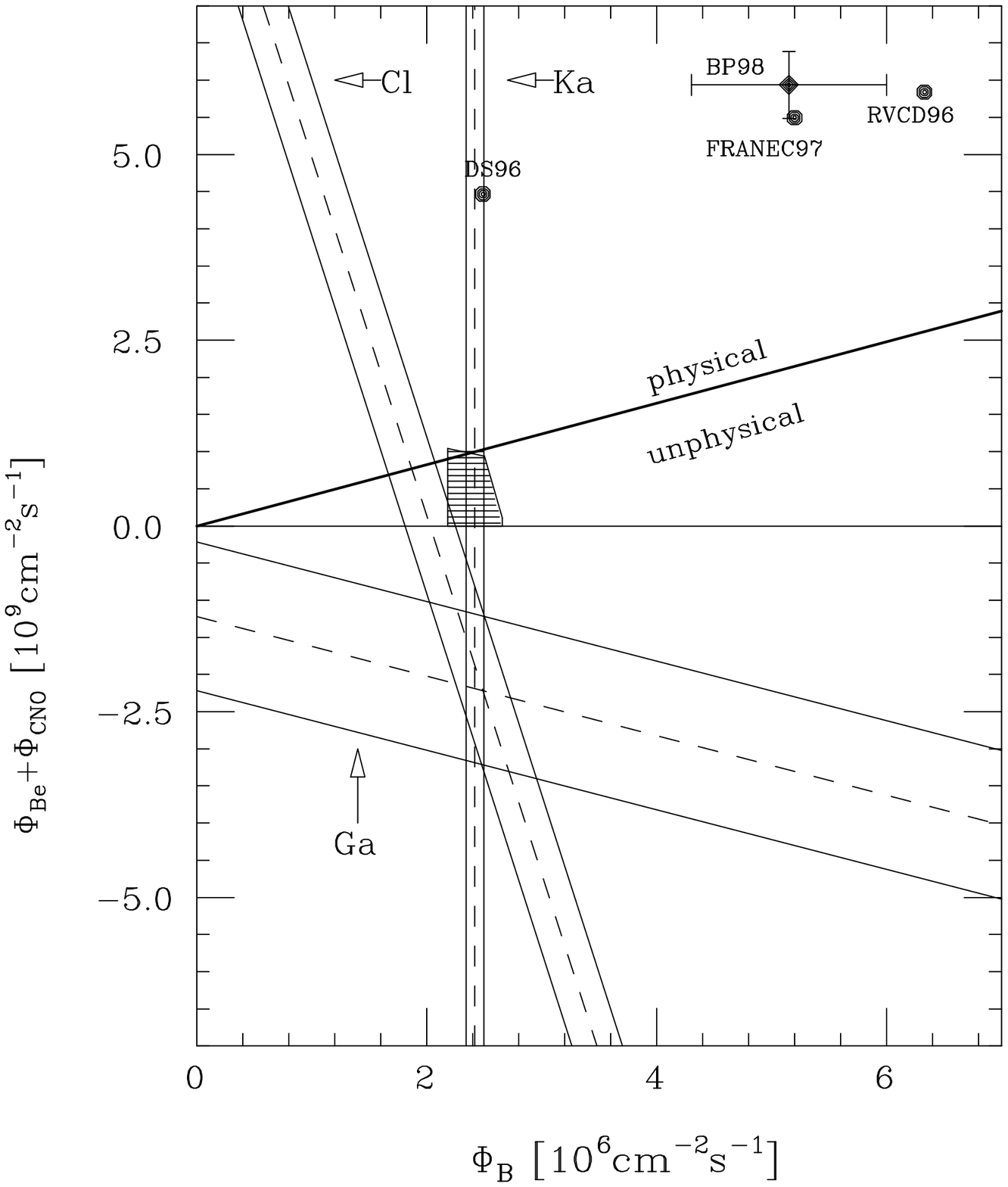,width=10cm}\hfill}
  \vspace*{0.05truein}		
  \fcaption{
The $^8B$ and $^7Be+CNO$ neutrino fluxes, 
consistent with the luminosity constraint
and experimental results for standard neutrinos.
The dashed (solid) lines correpond to the central ($\pm 1\sigma$) experimental
values for Chlorine, Gallium 
and $\nu - e$ scattering  experiments. 
The dashed area corresponds to the region within $3\sigma$ from each
experimental result.
The predictions of solar models including element diffusion (full circles)
~\cite{DS96,bp98,RCVD96,ciacio},
are also shown.
The thick diagonal line corresponds to the helioseismic lower
limit on $\Phi(Be)$, see text.
}
  \label{figbebo}
  \end{figure}

Since helioseismology determines $u=P/\rho$,
by using the hydrostatic equilibrium equation,
$dP/dR=-G M_R \rho /r^2$, one can determine
$\rho=\rho(u)$, i.e. the density profile
is also given by means of helioseismology.
Furthermore, by using the classical 
perfect gas law one has  $KT=u \mu$,
see the previous section, so that also the
temperature is given by helioseismology, {\it except
for the mean molecular weight $\mu$}.
As previously remarked, one can determine constraints
on $\mu$, which translate into constraints on 
the neutrino production rates.

This approach has been applied to the study of $hep$-neutrinos
in \cite{noihep}.
As well known,
the excess of highest energy solar-neutrino events observed by
Superkamiokande can be in principle explained by an anomalously high
$hep$-neutrino flux $\Phi_{\nu}(hep)$.
Without using SSM calculations, from the solar luminosity
constraint it was found  that $\Phi_\nu(hep)/S_{13}$ cannot exceed the SSM
estimate by more than a factor three. If one makes the
additional hypothesis that $hep$ neutrino  production
occurs where the $^3$He concentration is at equilibrium,
helioseismology gives an upper bound which is (less then) two times
the SSM prediction.
We argue that the anomalous $hep$-neutrino flux
of order of that observed by Superkamiokande cannot be explained by
astrophysics, but rather by a large production cross-section.

In ref. \cite{berillio} a lower limit  on the
Beryllium neutrino flux on earth was found,
$\Phi(Be)_{min} = 1\cdot 10^9$ cm$^{-2}$ s$^{-1}$,
in the absence of oscillations,
by using helioseismic data, the B-neutrino flux measured by
Superkamiokande  and the hydrogen abundance at the solar center $X_c$
predicted by Standard Solar Model (SSM) calculations.
We remark that this abundance is the only result of SSMs
needed for getting $\Phi(Be)_{min}$.
Lower bounds for the Gallium
signal, $G_{min}=(91 \pm 3) $ SNU, and for the Chlorine
signal, $C_{min}=(3.24\pm 0.14)$ SNU, have also been derived.
They are about $3\sigma$ above the
corresponding experimental values, $G_{exp}= (72\pm 6)$ SNU \cite{gallex,sage}
and $C_{exp}= (2.56\pm 0.22) $ SNU \cite{homestake}.

We remark that predictions for $X_c$ are very stable among
different (standard and non standard) solar models, see \cite{berillio}.
In fact $X_c$ is essentially an indicator of how much hydrogen
has been burnt so far. The stability of $X_c$ corresponds to
the fact that any solar model has to account for the same
present and time integrated solar luminosity.

In Fig. \ref{figbebo} we summarize the present situation
concerning solar neutrino experiments. Helioseismology,
when supplemented  with the hydrogen abundance at the solar center $X_c$
given by SSM, provides  the lower bound $\Phi(Be) \geq	 4 \cdot
10^2 \Phi(B)$ (thick diagonal line). One sees that the
region within three sigmas from each experiment is almost
completely out of the physical domain. 

Along similar lines, a more extensive analysis has been presented in 
\cite{berezinsky} where the stronger bound on 
  Beryllium neutrinos, $\Phi(Be) \geq 1.6 \cdot 10^9$ cm${-2}$ s$^{-1}$
was found.

  \begin{figure}
 {\hfill \epsfig{figure=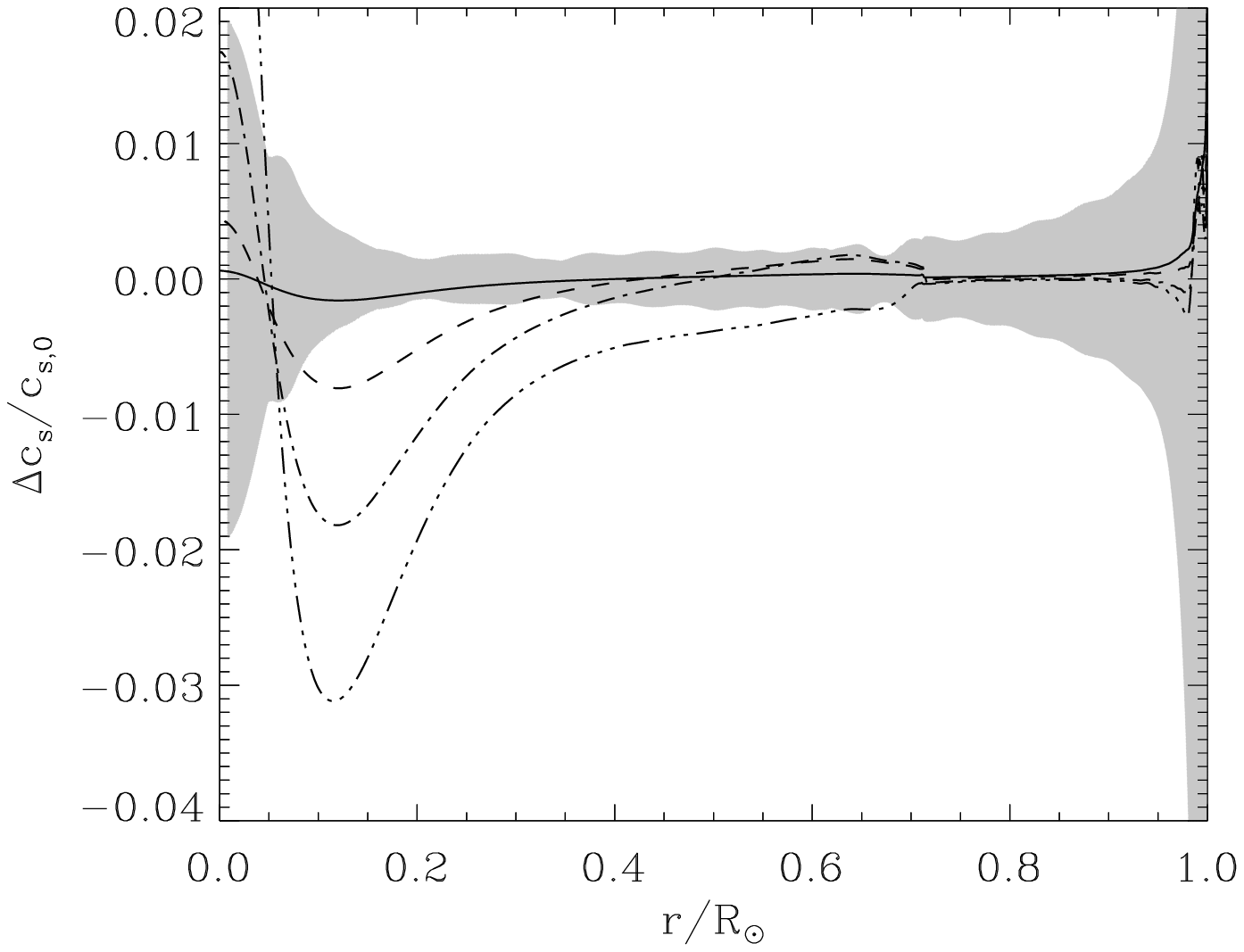,width=10cm}\hfill}
  \vspace*{0.2truein}		
  \fcaption{ Difference in sound-speed profiles of present-day solar
models with axion losses compared to the reference model in
the sense (Reference$-$Model)/Reference. 
Different line types correspond to different values of the
axion-photon coupling constant: $g_{10}$=4.5 (solid line), 10
(short-dashed), 15 (dash-dotted), 20 (dash-dot-dot-dotted.
The shaded area
reflects the ``$3\sigma$'' uncertainties in the infered sound speed of the
seismic model. From \cite{raffelt}}
  \label{figraffelt}
  \end{figure}

\section{Helioseismic tests of exotic solar models:axions from 
the sun?}

Helioseismology severely constrains possible
  deviations from standard solar models, allowing e.g. the derivation
 of new
  limits on anomalous solar energy losses.
In ref. \cite{raffelt} as an example of nonstandard energy loss channel,
the Primakoff conversion of photons in the Coulomb fields of charged particles,
$\gamma+Ze \rightarrow Ze+a$ has been considered.

Axion emission from the sun alters the hydrogen burning now and in the past.
More hydrogen is being burnt and consequently the central solar
temperature increases with respect to the SSM prediction. 
In addition more  hydrogen has been burnt
into helium in the past, so that the helium abundance in the solar center
differs from the SSM value.
All this affects the sound speed profile, see Fig. \ref{figraffelt},
and the photospheric helium
abundance.

 For an axion-photon coupling
  $g_{a\gamma}\lapprox 5\times10^{-10}~{\rm GeV}^{-1}$ the solar model is
  almost indistinguishable from the standard case, while
  $g_{a\gamma}\gapprox 10\times10^{-10}~{\rm GeV}^{-1}$ is probably
  excluded, corresponding to an axion luminosity of about
  $0.20\,L_\odot$.  This constraint on $g_{a\gamma}$ is much weaker
  than the well-known globular-cluster limit, but about a factor of 3
  more restrictive than previous solar limits \cite{raffelt}.

  \section{Acknowledgements}
We are grateful to V. Berezinsky and S. Degl'Innocenti for useful discussions,
and to  W.Dziembowski and R. Sienkiewicz for providing
us with  results of their solar code.

  \end{document}